\def\year{2021}\relax
\documentclass[letterpaper]{article} % DO NOT CHANGE THIS
\usepackage{aaai21}  % DO NOT CHANGE THIS
\usepackage{times}  % DO NOT CHANGE THIS
\usepackage{helvet} % DO NOT CHANGE THIS
\usepackage{courier}  % DO NOT CHANGE THIS
\usepackage[hyphens]{url}  % DO NOT CHANGE THIS
\usepackage{graphicx} % DO NOT CHANGE THIS
\usepackage{natbib}  % DO NOT CHANGE THIS AND DO NOT ADD ANY OPTIONS TO IT
\usepackage{caption} % DO NOT CHANGE THIS AND DO NOT ADD ANY OPTIONS TO IT
\usepackage[switch]{lineno}  %
\usepackage{nicefrac}
\usepackage{dsfont}

\usepackage{subcaption}

\usepackage{soul}
\usepackage{amsmath}
\usepackage{amsbsy}

\usepackage{color}
\usepackage[textsize=scriptsize]{todonotes}
\definecolor{blue}{RGB}{0, 93, 170}			%Go Big Blue!
\definecolor{darkgreen}{RGB}{0, 102, 0}
\definecolor{myorange}{RGB}{1,0.5,0}

\newcommand{\argmax}[1]{\underset{#1}{\operatorname{arg}\,\operatorname{max}}\;}

\renewcommand{\citet}[1]{\citeauthor{#1}~\shortcite{#1}}

\title{Modeling Voters in Multi-Winner Approval Voting}
\author {
        Jaelle Scheuerman,\textsuperscript{\rm 1}
        Jason Harman, \textsuperscript{\rm 2}
        Nicholas Mattei, \textsuperscript{\rm 1} 
        K. Brent Venable \textsuperscript{\rm 3}  \\
}
\affiliations {
    \textsuperscript{\rm 1} Tulane University, New Orleans, LA, USA \\
    \textsuperscript{\rm 2} Louisiana State University, Baton Rouge, LA, USA \\
    \textsuperscript{\rm 3} University of West Florida and Institute for Human and Machine Cognition, Pensacola, FL, USA \\
    jscheuer@tulane.edu, jharman@lsu.edu, 
    nsmattei@tulane.edu, bvenable@ihmc.us 
}

\begin{document}
%\linenumbers % For submission 

\maketitle

\begin{abstract}
In many real world situations, collective decisions are made using voting and, in scenarios such as committee or board elections, employing voting rules that return multiple winners. In multi-winner approval voting (AV), an agent submits a ballot consisting of approvals for as many candidates as they wish, and winners are chosen by tallying up the votes and choosing the top-$k$ candidates receiving the most approvals. In many scenarios, an agent may manipulate the ballot they submit in order to achieve a better outcome by voting in a way that does not reflect their true preferences. In complex and uncertain situations, agents may use heuristics instead of incurring the additional effort required to compute the manipulation which most favors them. In this paper, we examine voting behavior in single-winner and multi-winner approval voting scenarios with varying degrees of uncertainty using behavioral data obtained from Mechanical Turk. We find that people generally manipulate their vote to obtain a better outcome, but often do not identify the optimal manipulation. There are a number of predictive models of agent behavior in the COMSOC and psychology literature that are based on cognitively plausible heuristic strategies.  We show that the existing approaches do not adequately model real-world data. 
We propose a novel model that takes into account the size of the winning set and human cognitive constraints, and demonstrate that this model is more effective at capturing real-world behaviors in multi-winner approval voting scenarios. 
\end{abstract}

\section{Introduction}

Computational Social Choice (COMSOC) investigates computational issues surrounding the aggregation of individual preferences for the purposes of collective decision-making \cite{BCELP16a}. There is a rich body of work that focuses on the computational complexity of manipulating elections under various voting rules \cite{FaPr10a}. In some cases, it is easy for a voter to compute the optimal manipulation for a given scenario, e.g.,  an agent casting a ballot for their second ranked candidate, who would win over their third ranked candidate, when their most preferred candidate has no chance in a plurality election.

However, in cases when it is computationally hard for the agent to manipulate an election optimally, it is assumed that voters will report their true preferences rather than trying to strategize
%and submit an insincere report that may lead to higher overall utility, i.e., manipulate their vote 
\cite{faliszewski2010acm,FaPr10a}. Voting truthfully is just one possible heuristic that voters may use when faced with complex voting scenarios in which the optimal strategy is not easy to compute. %A recent study of voting behavior in multi-winner approval elections showed that the majority of voters did not vote truthfully or optimally \cite{scheuerman_heuristics_2020}. Instead, the predominant strategy used was to prioritize the highest utility candidates. 
For example, a recent study showed that in a plurality election where there was uncertainty over how many more votes were to be cast and a preferred candidate had no chance to win, voters would compromise and vote for the current leader \cite{TMG15a}.  In other work on allocations, agents have been observed using local strategies to manipulate when faced with a complex computation to find the optimal strategy \cite{MWPS15a}.

The effectiveness of a particular heuristic depends on the environment in which that heuristic is being used. Decision science research has examined heuristic decision making in complex and uncertain situations. Sometimes, heuristics are viewed as second best shortcuts, when the environment is too complex to use rational strategies \cite{tversky1974judgment}. However, researchers have also shown that heuristics work in the real world, leveraging natural cognitive abilities that exploit the structure of the environment, often leading to better outcomes with the use of less information. Key to this view of heuristics is that in uncertain environments, decisions made by ignoring some information can sometimes achieve better performance than more complicated optimization strategies \cite{gigerenzer1999fast}. In many situations, heuristics have been shown to outperform solutions that use more complex algorithms, e.g., stock market predictions \cite{demiguel2007optimal}. 

In this paper, we examine both the effectiveness of heuristics and the accuracy with which these heuristics model real-world decision making in a multi-agent setting, namely, in single-winner and multi-winner approval voting elections. We consider scenarios both with and without uncertainty that is represented, in our case, as missing ballots in the election.  Approval voting is a set of methods for aggregating group preferences that is particularly popular among economists, computer scientists, psychologists, and beyond \cite{LaSa10a,BrFi07c}.  There are even multiple political action committees (PACs) in the United States\footnote{\url{https://www.electionscience.org/}}
%e.g., The Center for Election Science\footnote{\url{https://www.electionscience.org/}}, 
committed to seeing the United States change voting procedures to approval voting since voters are allowed to express a preference over a set of candidates and not just a single one \cite{LaVa08a}.  
%In France, a large study was run parallel to the 2002 election, showing that many voters would have preferred approval ballots to traditional plurality ballots \cite{LaVa08a}.
%
%
%
%
%In approval voting, an agent may vote for as many candidates as they wish. Winners are chosen by tallying up the votes and choosing the top-$k$ candidates receiving the most votes. 
%
Multi-winner approval voting is interesting as a voter has potentially multiple sincere ballots they can cast, with some being better than others \cite{MPRZ08a,walsh2007uncertainty}. 
%Under the basic approval voting rule, an optimal manipulation, i.e., maximizing the voter's utility, can be computed in polynomial time when an agent has complete information about the preferences of all the voters and utilities are dichotomous \cite{MPRZ08a}. However, it has been shown that when information about voting preferences is missing, computing the possible winners or manipulating the vote is computationally difficult \cite{walsh2007uncertainty}. Although manipulation may be computationally hard, in some instances, voting truthfully or using other heuristics may still maximize a voter's expected utility. 
%

Understanding the behavioral component of strategic actions plays an important and nuanced role in agent models as the complexity of the voting scenarios increases. The ability to approve multiple candidates, elect multiple winners, and partial information on current votes are common in many real-life election settings. A computational model able to capture this contextual information and predict voter's choices in such settings, provides a better understanding of how voters behave at both the individual and group level. Prediction accuracy on real-world data is an important metric for evaluating models of how humans vote.  Accurate models play a fundamental role in providing more reliable forecasts, plausible explanations of voter behavior, and new models for complexity analysis of voting settings \cite{Mattei:Closing}.

\smallskip
\textbf{Contributions.}
We perform a novel experiment to look at voter behavior in multi-winner approval voting under uncertainty.  
%This behavioral experiment is the first one to look at a setting with uncertainty.  
We identify cognitively plausible heuristics from the psychology literature that may serve as models of voters' responses, and we provide a quantitative analysis of when and how often these models are deployed in the real world.
%characterize the responses of real people in these settings and quantify the strategies they employ.\nick{Make this more concrete!}.  
We evaluate these heuristics against others from the COMSOC literature on voting in terms of accuracy in predicting voters' behavior, showing that none of these existing models are very accurate. Finally, we propose a new model which exhibits a significant performance improvement by taking into account contextual information, such as the size of the winning set, as well as human cognitive constraints.
   
%We then evaluate mand proposed a novel model to capture the behavior of actual decision makers.  We find that our model is more predictive than models previously proposed. \nick{punch this up more when we have everything}.

\section{Approval Voting}
%We give a brief overview of the mathematical formalism used to study approval voting and formally define the heuristics that we will consider in this paper.

Following \cite{AGGM+15a} and \cite{Kilg10a} we consider a social choice setting $(N,C)$ where we are given a set $N = \{1, \ldots, n\}$ of voting agents
%\brent{I changed the candidates to 1...m instead of c_1 ...c_m to be consistent with the use of j as a candidate. Nick I hope this is not blasphemy for the AV folks....}
and a disjoint set $C = \{c_1, \ldots, c_m\}$ of candidates.  Each agent $i \in N$ expresses an approval ballot $A_i \subseteq C$ which  gives rise to a set of approval ballots $A = \{A_1, \ldots, A_n\}$, called a profile.  We study the multi-winner approval voting rule that takes as input an instance $(A, C, k)$ and returns a subset of candidates $W \subseteq C$ where $|W| = k$ called the winning set. Approval Voting (AV) finds the set $W \subseteq C$ where $|W| = k$ that maximizes the total weight of approvals (approval score), $AV(W) = \sum_{i\in N} |W \cap A_i|$. Informally, the winning set under AV is the set of candidates approved by most voters.

In some cases, it is necessary to use a tie-breaking rule in addition to a voting rule to enforce that the size of $W$ is indeed $k$. Tie-breaking is an important topic in COMSOC and can have significant effects on the complexity of manipulation of various rules even under idealized models \cite{AGM+13a,MNW14a,OEH11a}. Typically in the literature, a lexicographic tie-breaking rule is given as a fixed ordering over $C$, and the winners are selected in this order. However, in this paper, as discussed in \cite{AGM+13a}, we break ties by selecting a winner uniformly at random from the tied set. 
%to more closely simulate a real-world election.

\begin{table}[htbp]
\centering 
{\small
\begin{tabular}{|c|c|c|c|c|c|}
\hline 
\textbf{Candidates ($C$)}: & \textbf{A} & \textbf{B} & \textbf{C} & \textbf{D} & \textbf{E} \\
\hline
Utility ($u_i$): & 0.05 & 0.10 & 0.01 & 0.25 & 0 \\
\hline
\end{tabular}
}
\caption[Sample approval voting profile]{Example of the utility of a voter $i$ for 5 candidates A,B,C, D and E.}
\label{tab:sample-profile}
\end{table}

%In order to align our
Similarly to work in the literature on decision heuristics, e.g., \citet{gigerenzer1996reasoning} and COMSOC, e.g., \citet{MPRZ08a},
we assume that each agent $i \in N$ also has a real valued utility function $u_i: C \rightarrow \mathcal{R}$ (Table \ref{tab:sample-profile} shows an example with 5 candidates).  We also assume that the utility of agent $i$ for a particular set of winning candidates $W \subseteq C$ is $u_i(W)=\sum_{c \in W} u_i(c)$ (with slight abuse of notation). If $W$ is the subset elected by the voting rule we will refer to $u_i(W)$ as agent $i$'s utility for the \emph{outcome} of the election.

\subsection{Truthfulness in Approval Voting} \label{sec:truthfulness}

Studies of approval voting for multi-winner elections span nearly 40 years \cite{brams1980approval}.  For nearly that entire period, there has been an intense discussion of the strategic aspects of approval balloting \cite{brams1982strategic}.  Researchers over the years have made a variety of assumptions and (re)definitions of what makes a particular vote either \emph{truthful} or \emph{strategic} \cite{LaSa10a,Niem84a,brams1982strategic}.  According to \citet{brams1982strategic}: ''A voter votes sincerely if and only if whenever he votes for some candidate, he votes for all candidates preferred to that candidate.''  However, even this definition is debated as there can be multiple sincere strategies \cite{Niem84a}.  This definition is used in recent COMSOC literature to define \emph{Sincere Ballots} \citet{endriss2007vote}, of which there may be many for a given scenario. This is a subtle issue as when it is assumed that agents have dichotomous preferences, then multi-winner approval voting is incentive compatible since a  complete ballot, i.e., one for all candidates with positive utility, and a sincere ballot are the same.  However, if agents have linear preferences over the candidates, as they do in our settings, then there may be multiple sincere ballots that are not complete \cite{MPR08b,MPRZ08a}.

We make the following distinctions which are supported by both psychology research on heuristic strategies discussed in Section \ref{section:heuristics} and discussions in the COMSOC community about voting in approval voting scenarios.
\begin{description}
\item [Complete Ballot:] a voter submits a ballot approving all candidates for which they have positive utility.

\item [Sincere Ballot:] a voter submits a ballot in which if a voter approves a particular candidate, then she also approves all candidates that are preferred to that candidate \cite{endriss2007vote,MPRZ08a,brams1982strategic}. 
%Intuitively, this is an assumption of monotonicity over the preferences and captures many of the votes one would cast in the \emph{take the X best} heuristic.

%A \emph{sincere vote}, which includes the definitions of \citet{endriss2007vote}, \citet{MPRZ08a}, and \citet{brams1982strategic}, is one in which if a voter prefers a particular candidate, then he approves all candidates that are preferred to that particular candidates. Intuitively, this is an assumption of monotonicity over the preferences and captures many of the votes one would cast in the \emph{take the X best} heuristic.
\end{description}

As an example of a \emph{Complete Ballot} versus a \emph{Sincere Ballot}, consider a voter having the set of preferences $[A=0.4,B=0,C=0.2,D=0.01]$. Given these preferences, a \emph{Complete Ballot} is $[A,C,D]$, while a \emph{Sincere Ballot} could be either $[A]$, $[A,C]$, $[A,C,D]$,  or $[A,B,C,D]$.

%We argue and will use the terminology that any vote that is not completely truthful by our definition is considered strategic.  While it is the case that these votes may be sincere, we argue they are not completely truthful as, given the definitions above, it is strategically leaving some information out. In what follows, we consider, as does much of the literature, the question of which strategic vote to use, and what internal heuristics one may be using to decide it.

\section{Related Work}

%%% RESTORE LATER!!

%Approval voting is a set of methods for aggregating group preferences that is particularly popular among economists, computer scientists, psychologists, and beyond \cite{LaSa10a,BrFi07c}.  There are even multiple political action committees (PACs) in the United States, e.g., The Center for Election Science\footnote{\url{https://www.electionscience.org/}}, that are committed to seeing the United States change voting procedures to approval voting.  One reason for this popularity is the idea that participants are allowed to express a preference over a set of candidates and not just a single one.  In France, a large study was run parallel to the 2002 election, showing that many voters would have preferred approval ballots to traditional plurality ballots \cite{LaVa08a}.

%%%%

%Academic studies of approval voting (AV) include numerous studies about 
%Approval voting is a set of methods for aggregating group preferences that is particularly popular among economists, computer scientists, psychologists, and beyond \cite{LaSa10a,BrFi07c}.
The complexity of manipulation for various types of AV has received considerable attention in the COMSOC literature \cite{BCELP16a}.  If agents act rationally and have full information about the votes of other agents, when agents have \emph{Boolean utilities}
%, i.e., when all agents either have utility 1 or 0 for candidates they approve or disapprove of, respectively, 
AV is strategy-proof.  When agents have general utilities, finding a vote that maximizes the agent's utility can be computed in polynomial time \cite{MPR08b,MPRZ08a}.  For variants of AV, including Proportional AV, Satisfaction AV, and RAV, the complexity of finding utility-maximizing votes ranges in complexity from easy to coNP-complete \cite{AGGM+15a}.

%(see \citet{AGGM+15a} for a full discussion of the computational issues of these variants of approval voting).

Theoretical work in COMSOC often makes worst-case assumptions, e.g., that manipulators have complete information \cite{BCELP16a}. There are efforts to expand these worst-case assumptions to include the uncertain information and agents that are not perfectly rational, to more closely model the real-world \cite{Mattei:Closing}. In \citet{ReEn12a}, agent behaviors are measured when agents are given access to poll information. 
%and agent behaviors are modeled as being $k$-pragmatist, i.e., they only look at the $k$ highest ranked candidates when deciding whether or not to make a strategic decision. 
In \citet{MLR14a}, agents are modeled as behaving in \emph{locally dominant}, i.e., myopic ways.  
%i.e., they take into account only a small number of possible outcomes when deciding whether or not to act strategically in a particular voting setting.  
A survey of recent work on issues surrounding strategic voting is given by \cite{Meir18a}.

There is a growing effort to use simulations and real-world data to test various decision-making models, e.g., \cite{MaWa17,MaWa13a,Mattei:Closing}. Within the economics and psychology literature, there have been several studies of approval voting and the behavior of voters. Perhaps the most interesting and relevant to our work are the studies of \cite{RHT07a}, which focus on the elections of various professional societies where approval balloting.  \citet{RHT07a} find that many voters use a \emph{plurality heuristic} when voting in AV elections, i.e., they vote as if they are in a plurality election, selecting only their most preferred candidate.  \citet{ZMD15a} use approval voting data from Doodle and conclude that the type of poll has a significant effect on voter behavior. In both of these works, only AV with a single winner was investigated, and both works relied on real-world elections where it was not possible to tease out the relationship between the environment and decisions.

%was used and the work of \cite{ZMD15a} where many approval voting settings were obtained from Doodle, an online polling platform. In \cite{RHT07a} election data is used along with proposed heuristics for individual choice behavior, the conclusion is that many voters use a \emph{plurality heuristic} when voting in AV elections, i.e., they vote as if they are in a plurality election, selecting only their most preferred candidate.  In both of these works, only AV with a single winner was investigated, and both works relied on real-world elections where it was not possible to tease out the relationships between environment and decision. 

%To our knowledge, the work presented in this paper is the first that examines human voting behavior in multi-winner approval settings. The behavioral experiment presented here examines how voters select different decision-making heuristics, depending on the underlying environmental factors (i.e., number of winners and number of missing ballots), and this parameterization is also novel. %We also show how strategies, such as voting truthfully or for a single preferred candidate, can sometimes be effective in maximizing one's expected utility.
 
Three recent papers address strategic voting under the plurality rule, where agents are making decisions in uncertain environments.  First, \cite{TySc16a} studied the voting behavior of agents under the plurality rule with three options.  They find that the amount of information available to the voters affects the decision on whether or not to vote strategically and that in many cases, the strategic decisions do not affect the outcome of the plurality vote. Second, in \cite{TMG15a} an online system is presented where participants vote for cash payments in a number of settings using the plurality rule under uncertainty. 
%Two specific scenarios are studied: one where a user votes after being given access to a large pre-election poll and the second where agents vote simultaneously and can update their votes.  
They find that most participants do not engage in strategic voting unless there is a clear way to benefit. %In the iterative setting, 
Indeed, most voters were lazy, and if they did vote strategically, they would do a one-step look-ahead or perform the best response myopically. Finally, in \cite{FLMG19a}, a comprehensive study using both past datasets and newly collected ones examines the actual behavior of agents in multiple settings with uncertainty versus behavior that is predicted by a number of behavioral and heuristic models.  The paper proposes a novel model of user voting behavior in these uncertain settings called \emph{attainable utility}, where agents consider how much utility they would gain versus the likelihood of particular candidates winning given an uncertain poll.  They conclude that the attainable utility model is able to explain the behavior seen in the experimental studies better than existing models and even perform near the level of state-of-the-the-art machine learning algorithms in modeling users' actual behavior.

We expand upon these recent works on plurality to consider heuristics in the significantly more complex setting of multi-winner approval voting with uncertainty. We show that, in this context, simple heuristics are inadequate to capture voting behavior and applying the attainability-utility model directly results in limited predictive capabilities. 
We then extend the attainability-utility model 
to incorporate more information of the voting scenarios while maintaining cognitive plausibility, proving that this is key to greatly enhance predictive accuracy.
%cannot be dircetly applied, it cna be adapted to is expanded to include a threshold parameter, it can be used as an effective predictor of approval votes. 

%\st{it may be ecologically rational for voters to use heuristics over more complex optimization strategies}.

\section{Heuristics for Approval Voting}\label{section:heuristics}

We consider three heuristics inspired by the literature in cognitive science and COMSOC. These consist of \textit{Complete}, \textit{Take the X best}, and  \textit{Attainability-Utility (AU)}. We study these heuristics experimentally with behavioral data and evaluate them in terms of their predictive accuracy as a model of voter behavior. We show, in Section \ref{section:data} and  Section \ref{section:evaluation}, that though these strategies are frequently employed by real-world decision makers, none of them are able to forecast human votes with reasonable accuracy. To address this, we  present a novel heuristic model for multi-winner approval voting we call \emph{Attainability-Utility with Threshold (AUT)}.

\subsection{Cognitive Heuristics}

We start by considering two simple heuristics which ignore the current voting scenario information and use only the voter's utility for each candidate.  The data from our real-world experiment, described in Section \ref{section:data} and the Technical Appendix, shows that these heuristics are indeed adopted by voters in the context of single winner and multi-winner approval voting.  Examples of ballots corresponding to these heuristics are shown in Table \ref{tab:trivial-candidate} and \ref{tab:trivial-leader}.

\begin{description}
\item [Complete.] When an agent uses the \textit{Complete} heuristic, they approve all candidates for which they have positive utility. This corresponds to a Complete Ballot.

\item [Take the X best.] When an agent $i$ votes with the \textit{Take the X best} heuristic, they vote for a subset of the complete vote. First, they order the list of candidates by the utility value, $c_j>c_h$ when $u_i(c_j)>u_i(c_h)$.
%: $c_1 > ... > c_m$ where $u_i(j_1) > ... > u_i(j_m)$.
In this paper, we examine situations where the candidates' utility values are not tied, so no tie-breaking rule is required.  
%Formally, top x candidates $T = t_1 > ... > t_x$ where $u_1(t_1) > ... > u_X(t_X)$. 
The agent will then vote for the top-$X$ candidates in this ordering with $1 \leq X\leq m$. $X$ could be calculated using a magnitude cut off or a proportional difference between preferences \cite{brandstatter2006priority}, or setting $X$ to be the number of winners in the election ($X = k$). 
\end{description}

\noindent
In the example shown in Table \ref{tab:sample-profile}, the \textit{Complete} ballot is $\{A,B,C,D\}$ and the \textit{Take X best} with $X=2$ is $\{D,B\}$. 
Observe that, for every setting of $X$ in \emph{Take the X best} we have an instance of a \emph{Sincere Ballot} as described by \citet{brams1982strategic}.

%\subsection{Heuristics Based on Attainability and Utility}

%In this section we present two heuristics, one from the literature and one our novel extension, that consider the utility a candidate brings to a voter as well as the voters' perception of how likely a particular candidate is to win.

\subsection{Attainability-Utility (AU) Heuristic} \label{section:au}

\citet{FLMG19a} present a heuristic for plurality that explains the voting behavior of individual agents as a trade-off between attainability and utility. They define attainability as the likelihood that a candidate will win an election, given access to uncertain poll information about the current state of the election. The formula for computing attainability was first introduced by \citet{bowman_potential_2014} for voting over a set of binary issues and was then extended by \citet{FLMG19a} to the case of $m$ candidates. 

\begin{figure*}[ht]
\centering
\begin{minipage}[t]{0.31\linewidth}
	\centering 
    \includegraphics[width=\linewidth]{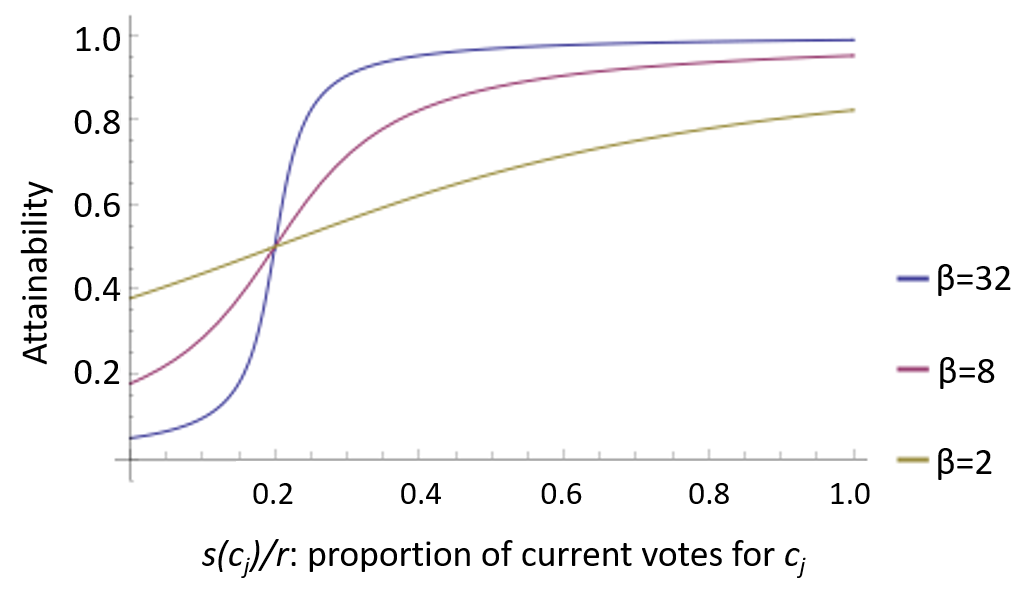}
    \caption{$A_{\beta}(c_j,s,r)$ for different values of $\beta$ in any single-winner approval election with five candidates.}
    \label{fig:betaparams}
\end{minipage}
\hspace{1em}
\begin{minipage}[t]{0.31\linewidth} 
    \centering 
    \includegraphics[width=\linewidth]{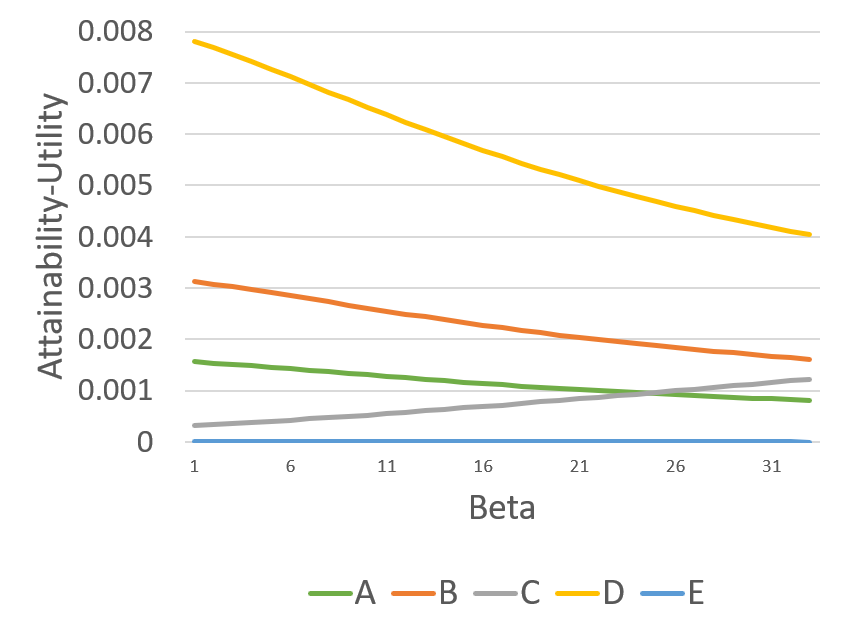}
    \caption{AU score of the five candidates from Scenario B for different values of $\beta$ when $\alpha = 1$.}
    \label{fig:autexample}
\end{minipage}
\hspace{1em}
\begin{minipage}[t]{0.31\linewidth} 
    \centering 
    \includegraphics[width=\linewidth]{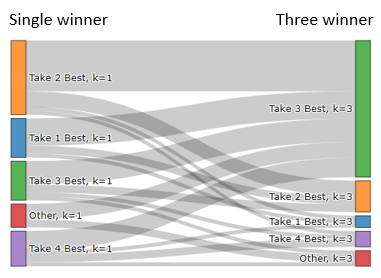}
    \caption{Visualization of the ballots cast by 54 subjects who voted in the single-winner and 3-winner elections with 0 missing votes.}
    \label{fig:sankey1-3}
\end{minipage}
\label{fig:all_graph}
\end{figure*}

Let us first review attainability in the case of plurality. Given a vector containing the current count of ballots in favor of each candidate, $s$, as well as the number of currently known ballots $r$,  the attainability of candidate $c_j$ is defined as:
%
%for all subsets of $m$ candidates in the power set of $C$ (i.e. $B = \mathcal{P(}$ $C$ $\mathcal{)}$).
%
%\begin{equation}\label{eq:attainability}
%A_\beta(j,s) = \frac{1}{\pi}\tan^{-1} \left( \beta %\left(\frac{s_j}{t} - v \right) \right) + 1/2
%\end{equation}
%
\begin{equation}\label{eq:attainability}
\resizebox{0.86\hsize}{!}{%
$A_\beta(c_j,s,r) = \frac{1}{\pi}\tan^{-1} \left( \beta \left(\frac{s(c_j)}{r} - \frac{1}{m} \right) \right) + \frac{1}{2}$,
}
\end{equation}

\noindent
where, $s(c_j)$ is the number of ballots for candidate $c_j$ and 
%and $t$ represents the total number of votes cast in the election (including any that are missing), $v$ represents the majority threshold required for an issue to be approved, and 
$\beta$ is a parameter modeling how a particular voter perceives candidate attainability.
%reflecting the perceived attainability of candidates by a voter. 
%Intuitively, given the amount of uncertainty in the election, $\beta$ is a voter dependent parameter that captures the voters' perception of how likely a candidate is to win the election. 
In Figure \ref{fig:betaparams} we depict three different ways in which the attainability values for a candidate vary with settings to $\beta$ in an election with 5 candidates as a function of the uncertainty remaining in the election. Lower values of $\beta$ mean the candidate’s attainability may seem more attainable when there is more uncertainty, with attainability increasing as more ballots are known. Higher values of $\beta$ scale the cotangent curve such that candidates may seem very unattainable when there is high uncertainty, but appear much more attainable as the uncertainty decreases.

We expand upon the work of \citet{FLMG19a} by (1) modifying AU to work in a multi-winner approval setting and (2) add a threshold parameter $\tau$ to model the minimum AU score that a candidate must have to be approved by the voter.

Translating the AU model to the multi-winner approval voting setting is non-trivial. AU in Equation \ref{eq:attainability} is only defined for single winners, and so we must modify the model to account for winning sets.

In the original model by \citet{bowman_potential_2014} where \nicefrac{1}{2} of the ballots are required to attain the passage of binary proposals, they used the quantity  $(\nicefrac{s(c_j)}{r}-\nicefrac{1}{2})$ to capture the difference between the proportion of the current ballots for a proposal and the score necessary to win.  In the extension to plurality elections by \citet{FLMG19a}, the $\nicefrac{1}{2}$ was changed to $\nicefrac{1}{m}$, to capture the idea that a candidate is more attainable if they are closer to having a plurality of the total current ballots, i.e., ($\nicefrac{s(c_j)}{r} - \nicefrac{1}{m}$). 

In both \citet{bowman_potential_2014} and \citet{FLMG19a}, the voter is intuitively comparing the proportion of current ballots for candidate $c_j$ to a uniform probability assumption over all possible outcomes. 

To apply this notion to winning sets of $k$ winners, we modify the $\nicefrac{1}{m}$ to be $\nicefrac{1}{(mk)}$.  Hence, when we have only one winner, $k=1$, we recover the model of \citet{FLMG19a}, and as the number of winners grows, the denominator $mk$ becomes larger, requiring $c_j$ to have a smaller percentage of ballots to appear attainable. Intuitively, this represents the idea that a desired candidate is more attainable as the size of the winning set increases as they need fewer total ballots cast.
%
%The component $\frac{1}{m}$ denotes the percentage of votes that candidate j must have to be attainable. In the plurality setting discussed in \citet{FLMG19a}, $m$ was used to represent the number of candidates in an election. We adapt the formula above to multi-winner approval elections, by replacing $\frac{1}{m}$ with $\frac{1}{{m \choose k}}$ where $m$ is the number of candidates and $k$ is the number of winners in the election. In fact ${m \choose k}$ represents the number of possible winning subsets. As the number of winners increase, the denominator ${m \choose k}$ becomes larger, requiring $j$ to have a smaller percentage of the votes to appear attainable. Intuitively, this represents the idea that it is more attainable to elect a desired candidate as the size of the winning set increases.
%
%In Figure \ref{fig:betaparams} we depict three different ways in which the attainability values for a candidate vary with settings to $\beta$ in an election with 5 candidates as a function of the uncertainty remaining in the election.

% \begin{figure}[htbp]
%     \centering 
%     \includegraphics[width=0.45\textwidth]{Figures/betaparams.png}
%     \caption{$A_{\beta}(c_j,s,r)$ for different values of $\beta$ in any single-winner approval election with five candidates.}
%     \label{fig:betaparams}
% \end{figure}

%In a plurality election, uncertainty is accounted for in the attainability equation by setting $t$ to the total expected number of votes in the election. This is not sufficient 
%Todo: explain the heuristic about why more candidates is more attainable

In the plurality setting, the number of current ballots $r$ coincides with the current total number of approvals.  This is, of course, not the case in approval voting, where people are allowed to approve up to $m$ candidates. Thus, in our setting, $r$ represents the total number of approvals contained in all ballots that have been currently counted. For example, if we consider an example with five candidates, $m=5$, $A,B, C, D$, and $E$, and $n=5$ voters where two voters have submitted their ballots consisting of $\{A\}$ and $\{A,B,C\}$, we would have $r=4$.  We calculate the attainability in an approval election with missing ballots to be the mean attainability over all possible ways in which the remaining ballots can be cast.  We model this by considering the set of all  possible total approval counts in the election: $T = \{ t ~|~ r \leq t \leq r + \hat{n} m$\},  where $r$ is the number of approvals made so far, $\hat{n}$ is the number of missing ballots and  and $\hat{n}m$ represents the number of total possible approvals.
In the running example, $\hat{n}=3$, $\hat{n}m=15$, and $T=\{4,5,6,7, \ldots, 19\}.$
%when there are $\hat{n}$ missing ballots. 
%We calculate the attainability in an approval election to be the mean attainability over all possible values of $t$. 
We can now define the attainability of candidate $c_j$ in a multi-winner approval voting setting as:%\jaelle{Is this better?}
%
%*\nick{Jason says we throw in a sentence in the discussion that points out that there may be other ways to do this which underscores the issues with generalizing the model.}
%
\begin{equation} \label{eq:au-multiwinner}
\resizebox{0.86\hsize}{!}{%
    $A_{\beta}(c_j,s,T)= \frac{\sum_{t \in T} \frac{1}{\pi} tan^{-1} \left(  \beta \left( \frac{s(c_j)}{t} - \frac{1}{{mk}} \right) \right) + \frac{1}{2}}{|T|}$.%
}
\end{equation}

The utility of a subset of candidates $u(b), b \in \mathcal{P}(C)$ is the sum of the utilities of each candidate $c_j$ in $b$. Formally, with a slight abuse of notation, $ u(b) =\sum_{c_j \in b} u(c_j)$. Slightly abusing notation, we define the attainability of a set of candidates $b$, as the product of $A_\beta(c_j,s,T)$ for all $c_j \in b$ and $1-A_\beta(c_j,s,T)$ for all $c_j \notin b$, that is:

\begin{equation}
\resizebox{0.86\hsize}{!}{%
  $A_\beta(b,s,T)= \prod_{c_j \in b}   A_\beta(c_j,s,T) \cdot \prod_{c_j \notin b}   (1-A_\beta(c_j,s,T))$.%
}
\end{equation}

%
%When using the AU heuristic and given current voting scenario $s$, the voter approves $M^{AT}(s)$, i.e., the set of candidates $b$ that maximize the product of its attainability and utility.
%
%\begin{equation}\label{eq:at}
%M^{AT}(s) = \argmax{b \in \mathcal{P}(C)} A_\beta(b,s) \cdot u(b)
%\end{equation}
%
We then find a candidate subset $b$ that maximize the trade-off between attainability and utility by incorporating an additional parameter $\alpha \in [0,2]$: 
\begin{equation} \label{eq:au}
\resizebox{0.86\hsize}{!}{%
$M^{AU}_{\alpha,\beta}(s,u,T)= \argmax{b \in \mathcal{P}(C)} \left( (\epsilon + u(b))^{\alpha} \cdot (A_{\beta}(b,s,T))^{2-\alpha} \right)$.%
}
\end{equation}

%Note that we slightly abuse notation by sending in a set $b$ instead of a single $c_j$
%\begin{align} \label{eq:au}
%M^{AU}_{\alpha,\beta}(s,u,T) & = \argmax{b \in \mathcal{P}(C)}  (\epsilon + u(b))^{\alpha} \cdot \\
%& \{ \mathds{1}_{\{c_j \in b\}} \cdot A_{\beta}(c_j,s,T)^{2-\alpha} + \\
%& \mathds{1}_{\{c_j \notin b\}} \cdot (1 - A_{\beta}(c_j,s,T)^{2-\alpha}) \}
%\end{align}

Note that, when $\alpha$ is 0 then the voter only considers attainability, but when $\alpha$ is 2 she only considers utility. A small constant value, $\epsilon$, is also added to account for situations where the left hand component of the equation would result in $0^0$, i.e. when candidates have 0 utility and $\alpha = 0$. 
%Thus, given parameters $\alpha$, $\beta$, $\epsilon$ and a poll $s$, a voter using the AU heuristic will approve the following set of candidates:

%\jaelle{Is this a good justification for epsilon?}

%\subsubsection{Attainability-Utility in Approval Elections}

In \citet{FLMG19a}, voting behavior was predicted for a plurality election, where voters may approve of only one candidate at a time. %. In plurality elections, voters cast a ballot for a single candidate and the candidate with the ballots wins the election. 
In multi-winner AV, voters approve as many candidates as they wish, and the top $k$ candidates win the election. When the AU model is applied in the approval setting, voters cast the ballot described by $M^{AU}_{\alpha, \beta}$ as defined in Equation \ref{eq:au}.
%of candidates in $b$ that maximize the AU score. 
%When $\alpha = 2$, this model predicts ballot $b$ with the highest total utility. When $\alpha = 0$, it predicts the ballot with the highest attainability. 
Table \ref{tab:approval-au} shows which ballot is selected by $M^{AU}_{\alpha,\beta}$ for various settings of $\alpha \in \{0,1,2\}$ and $\beta \in \{1,2,\dots,32\}$ on the single winner approval voting election scenario ($s,u,T$) when $s$ and $u$ are described as in Table \ref{tab:trivial-candidate} and $T = \{r\}$ (that is, there are no additional missing ballots). These $\beta$ values were chosen to represent the broad range of shapes that the attainability function can represent as shown in Figure \ref{fig:betaparams}.

\begin{table}[h]
\centering
\resizebox{0.70\hsize}{!}{%
\begin{tabular}{| p{5em} | p{3.8em} p{3.8em} p{3.8em}|} 
 \hline
  & \multicolumn{3}{c|}{ \pmb{$\alpha$}} \\
 \hline
 \pmb{$M^{AU}_{\alpha,\beta}$} \textbf{ballot} & \textbf{0} & \textbf{1} & \textbf{2} \\
 \hline \hline
 D & $[1..32]$ & & \\
 \hline
 A,B,C,D,E & & $[1]$ & \\
 \hline
 A,B,D,E & & $[2..8]$ & \\
 \hline
 D,E & & $[9..21]$ & \\
 \hline
 B,D,E & & $[22..32]$ & \\
 \hline
 A,B,C,E & & & $[1..32]$ \\
 \hline
\end{tabular}%
}
\caption{Ballots predicted for Scenario A using the attainability-utility model (first column) and corresponding $\beta$ ranges (all other columns)
for $\alpha \in \{0,1,2\}$.} 
%and $\beta \in \{1,2,\dots,32\}$.}
\label{tab:approval-au}
\end{table}

Observing Table \ref{tab:approval-au} we note how, in the AU model, $\alpha = 0$ predicts a ballot containing only the leading candidate, while $\alpha = 2$ predicts the complete ballot of all candidates with positive utility. When $\alpha = 1$, different ballots are chosen, depending on the value of $\beta$. Behavior that prioritizes higher utility candidates without voting for leading candidates is not predicted by this model. 

It is worth noting that the prediction space for Scenario A, shown in Table \ref{tab:approval-au}, predicts only a single Take the X Best strategy (A,B,C,E) and does not predict the optimal ballot ($E$ for 1 winner, 0 missing votes as seen in Table \ref{tab:trivial-optimal}). This is counter to the observed behavior where most people used different Take the X Best strategies (see Figure \ref{fig:sankey1-3}). Hence, it is clear that when moving into an approval voting setting, the AU heuristic is inadequate as a model of voter behavior.

%\subsection{A Novel Model for Multi-winner Approval Voting} 
\subsection{The Attainability-Utility with Threshold (AUT) Heuristic}
\label{section:aut}

The trade-offs between attainability and utility proposed by \citet{FLMG19a} are common among cognitive models of decision making \cite{tversky1974judgment,gonzalez2002making} and are cognitively plausible for settings involving a single winner.
%As can be seen in Formula \ref{eq:au} 
When applied to multi-winner contexts, the formalization of the trade-offs entails calculating a value for every possible missing  ballot (see Formula \ref{eq:au}) which becomes increasingly implausible and cognitively demanding as $m$ increases. 
Hence, we introduce a new  model which instead assumes that participants calculate the AU value for each individual candidate, and then choose those candidates that surpass a threshold. This change makes the required mental computations cognitively plausible and is consistent with the adaptive toolbox theory of human decision making \cite{payne1993adaptive,gigerenzer2002bounded}, that people have multiple decision strategies that they choose to match to the environmental constraints of a decision scenario.

More formally, rather than considering the attainability-utility score (AU score) of each $b \in \mathcal{P}(C)$, we calculate the AU score for each individual candidate $c_j$:
\begin{equation} \label{eq:au-individual}
\resizebox{0.86\hsize}{!}{%
$AU_{\alpha,\beta}(c_j,s,u,T)= 
(\epsilon + u(c_j))^{\alpha} \cdot (A_{\beta}(c_j,s,T))^{2-\alpha}$.%
}
\end{equation}

%for short $A_\beta(c_j,s,T)$ for each individual candidate $c_j \in C$, using equation \ref{eq:au-multiwinner}, 
We introduce a threshold $\tau$ to set the minimum AU score that a voter that will lead to a voter approving $c_j$. The introduction of the threshold to the model was inspired by concepts in cognitive psychology \cite{busemeyer1993decision}. Decision thresholds are parameters that vary predictably based on factors in the environment (such as uncertainty). We expected that using thresholds in our AUT model would outperform AU because it more closely models the cognitive process used to vote in the current paradigm. In approval voting scenarios, the model predicts that a voter will submit a ballot approving of all candidates with an AU score greater than $\tau$. We will refer to this heuristic as \emph{ attainability-utility with threshold (AUT)}, and define the corresponding ballot for threshold $\tau$ as:
\begin{equation} \label{eq:aut}
\resizebox{0.86\hsize}{!}{%
$M^{AUT}_{\alpha,\beta}(s, u, T, \tau) = \{ c_j \in C
| \left( (\epsilon + u(c_j))^{\alpha} \cdot (A_{\beta}(c_j,s,T))^{2-\alpha} \right) \geq \tau \}$.%
}
\end{equation}
 
%Using this model we can simulate how a voter might rank their candidates by AU score and then choose to vote for only the candidates that exceed $\tau$. 
Figure \ref{fig:autexample} shows the AU score of each of the five candidates in Scenario B for different values of $\beta$ when $\alpha = 1$. We can see that if $ \beta = 2$, when $\tau = 0.007$, and $T = \{r\}$ the model predicts voting for only $D$, while when $\tau = 0.001$, then the model predicts voting for $\{A,B,D\}$. Note that for low values of beta, $\{A,B,D\}$ can be predicted by $\tau = 0.001$, while high values of beta will lead to a prediction of $\{A,B,C\}$. This exemplifies the flexibility AUT has in generating different ballots.
%will Hence, unlike the AU heuristic, the AUT heuristic, when used as a model of voter behavior, can predict the optimal strategy by selecting the optimal number of leading candidates in an approval voting election, given the appropriate settings for $\beta$ and $\tau$.
Of special note, 
%AU is a special case of AUT. 
is that in single-winner plurality elections where voters can only vote for one candidate, the AUT model simplifies to the AU model, accounting for human voting behavior in both plurality and approval voting.

% \begin{figure}[htbp]
%     \centering 
%     \includegraphics[width=0.45\textwidth]{Figures/autexample.png}
%     \caption{AU score of the five candidates from Scenario B for different values of $\beta$ when $\alpha = 1$.}
%     \label{fig:autexample}
% \end{figure}

%Todo: show how this model can predict optimal 

\section{Behavioral Data}\label{section:data}
%We ran a 
Our behavioral study aimed at investigating approval voting heuristics and included 104 participants recruited through Mechanical Turk.
%to participate in a voting heuristics study. 
Participants were asked to cast ballots in a voting game where they were presented with a number of multi-winner approval voting scenarios with a monetary value that was paid out when certain candidates won. Participants were paid \$1.00 to participate in the study and received a bonus of no more than \$8.00, which was determined by the outcome of hypothetical elections. All participants voted in the single winner scenarios (n=104). Participants were then randomly assigned to be part of a 2-winner (n=50) or 3-winner (n=54) election for the remainder of the study.  More information about this study can be found in the Technical Appendix.
%
%We consider each of the above heuristics in the context of data collected by \cite{scheuerman_heuristics_2020}. 
%Using this data set, we show how effective they are for predicting voting behavior.

In this analysis, we consider two scenarios in particular, Scenario A and Scenario B, shown in detail in Tables \ref{tab:trivial-candidate} and \ref{tab:trivial-leader}. In these scenarios, the participant faces a situation where each candidate generates a different amount of utility $u \in {0,0.1,0.05,0.10,0.25}$, paid as a reward if that candidate is in the winning set. In both scenarios, none of the participant's high utility candidates are leading the election, but all are within 1 approval of being tied for the lead. We observe how people respond, particularly considering if they cast a \emph{Complete Ballot} or a \emph{Sincere Ballot}, and how many sincere candidates they choose to approve.  The experimental data consists of responses to Scenarios A and B in 9 different conditions: where there are 1, 2 and 3-winners with 0, 1 or 3 missing ballots.  Hence we can examine how varying both the number of winners as well as the amount of uncertainty affects voter behavior.

\begin{table}[t!]
\centering 
{\small
\begin{tabular}{|c|c|c|c|c|c|}
\hline 
\textbf{Candidate}: & \textbf{A} & \textbf{B} & \textbf{C} & \textbf{D} & \textbf{E} \\
\hline
Utility: & 0.05 & 0.10 & 0.01 & 0 & 0.25 \\
\# Votes: & 3 & 3 & 3 & 4 & 3 \\
\hline
\end{tabular}
}
\caption{Scenario A design, including candidates, utilities and votes. \textit{Heuristic votes}: Complete: [A,B,C,E], Take 1 Best: [E], Take 2 Best: [E,B], Take 3 Best: [E,B,A]}
\label{tab:trivial-candidate}
\end{table}

\begin{table}[t!]
\centering 
{\small
\begin{tabular}{|c|c|c|c|c|c|}
\hline 
\textbf{Candidate}: & \textbf{A} & \textbf{B} & \textbf{C} & \textbf{D} & \textbf{E} \\
\hline
Utility: & 0.05 & 0.10 & 0.01 & 0.25 & 0 \\
\# Votes: & 3 & 3 & 4 & 3 & 3 \\
\hline
\end{tabular}
}
\caption{Scenario B design, including candidates, utilities and votes. \textit{Heuristic votes}: Complete: [A,B,C,D], Take 1 Best: [D], Take 2 Best: [D,B], Take 3 Best: [D,B,A]}
\label{tab:trivial-leader}
\end{table} 

Using $\chi^2$ analysis, we examined the responses from both scenarios across all conditions and found no significant difference in the distribution of responses between each scenario.  This means that even though the values of the candidates and the current leader changed, the voters behaved largely the same in both Scenarios A and B. Within each scenario, there was also no significant difference in how people voted as the number of missing ballots increased. However, significant differences ($P < 0.005$) were found when comparing the strategies used by voters in those conditions electing one or two winners compared to those electing three winners. In general, when voting in single-winner and 2-winner elections, participants cast a ballot for 2 or 3 candidates (single-winner: 57.9\%, 2-winner: 70.7\%) more often than other strategies. When participants voted in the 3-winner election, they usually cast a ballot for 3 candidates (61.7\%). Figure  \ref{fig:sankey1-3} shows how an individual voter's ballots changed as the number of winners increased from 1 to 3.

% \begin{figure}[htbp]
%     \centering 
%     \includegraphics[width=0.45\textwidth]{Figures/sankey-alluvial-trivial-leader-0-missing-ballots-1-3-winners.PNG}
%     \caption{Visualization of the ballots cast by 54 subjects who voted in the single-winner and 3-winner elections with 0 missing votes.}
%     \label{fig:sankey1-3}
% \end{figure} 

%In terms of employing a hueristic from Section \ref{section:heuristics}, 
We found that, in general, the majority of people vote sincerely with a \emph{Take the X Best} strategy: 78.8\% in Scenario A and 77.8\% in Scenario B. However, the value of X used by the voters changed depending on the individual and the size of the winning set, as seen in Figure \ref{fig:sankey1-3}. 
%The following three models aim to predict voter behavior by choosing a good value for X that represents the top X candidates the voter will

\section{Evaluation of Heuristics as Models}\label{section:evaluation}

Using our experimental data, we evaluated five approaches for predicting voter behavior over single-winner, two-winner, and three-winner conditions. First, we consider the effectiveness of a model that predicts optimal votes, i.e., the one that always maximizes the expected utility. We evaluate four other predictive models, each of which corresponds to the heuristics described in Section \ref{section:heuristics}, namely, \emph{Complete}, \emph{Take X Best} with X equal to the size $k$ of the winning set, AU and AUT.
%second approach applies the attainability-utility heuristic described above to determine if people consider the attainability and utility of each ballot. The final three approaches are based on the idea of the Take the X best heuristic, where people vote for their top X candidates. 
%First we consider if people vote for all candidates with positive utility. The next sets the value of X to be the size of the winning set. The third chooses the X candidates to vote for based on our novel model of the attainability-utility with threshold.

\subsection{Optimal Baseline}
As a baseline, we assume that people vote optimally, approving the ballot that maximizes their expected utility, which varies with the number of winners and missing ballots.  
%We first generate the set of all possible ballots that a voter could cast over $C$, i.e., power set of candidates $V = \mathcal{P}(C)$. 
Given the number of winners $k\in\{1,2,3\}$ and the number of missing ballots $\hat{n} \in \{0,1,3\}$
the expected utility $E$ of ballot $b \in \mathcal{P}(C)$ is: 
%given each combination of $k$ winners, and $n$ missing ballots.
$E(b,k,\hat{n}) = \sum_{c_j \in b} p(c_j,k,\hat{n}) u(c_j)$.
%\[\forall v \in V,~k \in \{1,2,3\},~n \in \{0,1,3\}\]
Assuming that all potential missing ballots are equally likely, $p(c_j,k,n)$ refers to the probability that candidate $c_j$ is elected, given that there are $k$ possible winners, $\hat{n}$ missing votes and that ties are broken uniformly at random. 

For each possible number of winners and the number of missing ballots, the Optimal Baseline, selects the ballot maximizing the expected utility:
\begin{equation}
\resizebox{0.55\hsize}{!}{%
$M^{Opt}(k,\hat{n}) = \argmax{b \in \mathcal{P}(C)} E(b,k,\hat{n})$.%
}
\end{equation}

The optimal ballots and the corresponding maximum expected values for Scenarios A and B can be found in Table \ref{tab:trivial-optimal}. Note that the Optimal Baseline in each of these scenarios and conditions corresponds to a variant of the \emph{Sincere Ballot} of  \textit{Take the X best}.
%The optimal ballots correspond to a sincere \textit{take the X best} strategy with the same value for $X$ in the corresponding Scenario A and B conditions \nick{don't we mean all versions of sincere ballots? there are words collioding here and htis is unclear}. 
For example, in both scenarios, the \emph{Take the 1 Best} heuristic is optimal  when there is a single winner and no missing ballots. However, it is optimal to \emph{Take the 2 Best} with 3 winners and no missing ballots.

\begin{table}[t!]
\centering
\resizebox{0.60\hsize}{!}{%
\begin{tabular}{| p{1em} | p{3.6em} p{3.6em} p{3.6em}|} 
 \hline
  & \multicolumn{3}{c|}{ \# winners ($k$)} \\
 \hline
 $n$ & 1 & 2 & 3 \\
 \hline
 0 &  \textbf{Take 1} \newline A: 0.12 \newline B: 0.13 & \textbf{Take 1} \newline A: 0.22 \newline B: 0.26 & \textbf{Take 2} \newline A: 0.31 \newline B: 0.36 \\
 \hline
 1 &  \textbf{Take 1} \newline A: 0.11 \newline B: 0.12 & \textbf{Take 2} \newline A: 0.21 \newline B: 0.22 &  \textbf{Take 2} \newline A: 0.30 \newline B: 0.31 \\
 \hline
 3 &  \textbf{Take 1} \newline A: 0.11 \newline B: 0.11 &  \textbf{Take 2} \newline A: 0.20 \newline B: 0.21 & \textbf{Take 2} \newline A: 0.29 \newline B: 0.29 \\
 \hline
\end{tabular}%
}
\caption{The maximum expected utility for Scenarios A and B, and the heuristic that achieves this optimal outcome. $n$ represents the number of missing ballots. }
\label{tab:trivial-optimal}
\end{table}

\subsection{Fitting Heuristics to Data}

In addition to the Optimal baseline, we fit each of the four heuristics to the data and tested their accuracy as predictive models of voter behavior.  In addition to testing \textit{Complete} and \textit{Take the $X$ Best}, where $X$ is set to be the number of winners $X = k$, we trained \textit{AU} and \textit{AUT} on the data collected from Scenarios A and B for each individual. The parameters and ranges we considered are: 

\begin{description}

%\item [Complete.] Here we model the \textit{Complete} heuristic as described in Section \ref{section:heuristics}, simulating a voter approving all candidates with positive utility.

%\item [Take the K Best.] As we saw in Section \ref{section:data}, voters employed various values of $X$ depending on scenario. Experimental data indicated that voters tend to vote for more candidates as the size of the winning set increases, so as a baseline, we predicted that people would vote for their top-k candidates, where $k$ refers to the number of winners. 

\item[Attainability-Utility (AU).] Using a grid search, we fit $\alpha \in \{0,1,2\}$ and $\beta \in \{1,2,\dots,32\}$.  

\item[Attainability-Utility with Threshold (AUT).] Since the behavioral results did not show a tendency to vote based on attainability alone, i.e., people rarely voted for only leading candidates with no utility or low utility, we choose to set $\alpha = 1$ and fit only $\beta$ and $\tau$. Using a grid search, we tested values for $\beta \in \{1,2,\dots,32\}$ and $\tau \in \{0,0.0005,\dots,0.10 \}$.
\end{description}

\begin{figure}[ht]
    \centering 
    \includegraphics[width=0.35\textwidth]{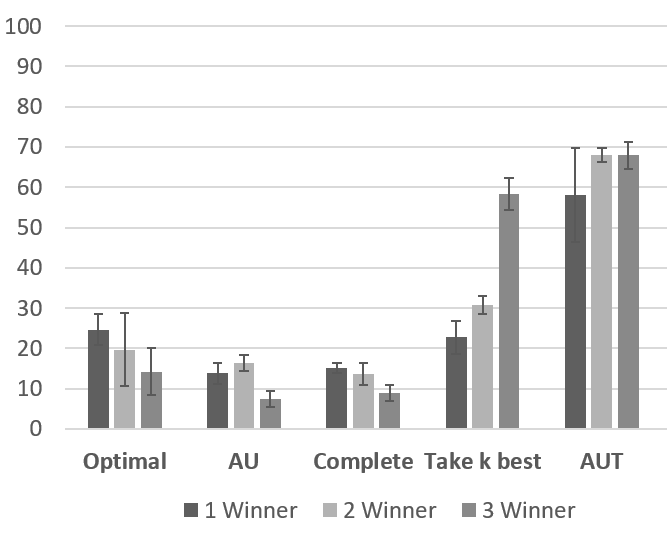}
    \caption{Mean and standard deviation of prediction accuracy for each model across conditions.}
    \label{fig:model-evaluation}
\end{figure}

We train the parameters of AU and AUT as follows. For each individual voter, when we fix $k$, we have 6 conditions to use for training and testing our model.  These 6 conditions correspond to the number of possible missing ballots, $\hat{n} \in \{0,1,3\}$, for Scenarios A and B. We use five of these observations to train the parameters of the AU or AUT model and predict the sixth.  Using a leave-one-out methodology, we do this for all possible splits of the data. We compute the accuracy over these six splits for each individual.  We then average this accuracy over all individual voters for each of the $k \in \{1,2,3\}$ winning set size conditions.  Each of the models were evaluated for their accuracy in predicting voter behavior in Scenarios A and B. The \emph{Optimal}, \emph{Complete}, and \emph{Take k Best} models are deterministic, depending on the scenario, the number of winners and the number of missing ballots.  The average prediction accuracy and standard deviation over the six responses for each of the winning set size conditions are reported in Table \ref{tab:model-evaluation} and shown in Figure \ref{fig:model-evaluation}.

\subsection{Evaluation Results and Discussion}

The model evaluation results show that a model of optimal behavior using expected utilities is not a good representation of voter behavior and supports the idea that people do not take the time to perform the calculations necessary to identify optimal solutions. We also found that people tend to not vote a \emph{Complete Ballot}. The experimental data indicated that voters tend to approve more candidates as the size of the winning set increases. Thus, we conjectured that the \emph{Take $k$ Best} model, predicting that people would vote for the number of top candidates that was equal to the number of winners ($k$) would perform well. However, this was refuted by our experimental results which showed   
%We found that this also did not perform well, with 
only a modest improvement in performance in the 3-winner condition. 

\begin{table}[ht]
\centering
\resizebox{0.95\hsize}{!}{%
\begin{tabular}{|c|c|c|c|}
\hline 
  & \textbf{1 winner} & \textbf{2 winner} & \textbf{3 winner} \\
\hline
Optimal: & 24.7\% (3.9\%) & 19.7\% (9.1\%) & 14.2\% (5.8\%) \\
AU: & 13.8\% (2.6\%) & 16.3\% (2.0\%) & 7.4\% (2.0\%) \\
Complete: & 15.1\% (1.2\%) & 13.7\% (2.7\%) & 9.0\% (2.0\%) \\
Take $k$ best: & 22.8\% (4.1\%) & 30.7\% (2.2\%) & 58.3\% (4.0\%) \\
AUT: & \textbf{58.1\% (11.6\%)} & \textbf{68.0\% (1.8\%)} & \textbf{67.9\% (3.4\%)} \\
\hline
\end{tabular}%
}
\caption{Mean prediction accuracy (standard deviation) for each model across conditions.}
\label{tab:model-evaluation}
\end{table}

Next, we tested whether or not voters make a trade-off between the utility and attainability of the candidates.  We first tried this by using the attainability-utility heuristic (AU), which considers the trade-off of attainability and utility for all subsets of candidates. We found that this also does not describe the voting behavior in multi-winner approval voting election. The poor performance of the AU model can likely be attributed to the way that it calculates the attainability-utility for every possible subset of candidates, rather than each individual candidate. The attainability-utility of ballots containing all utility generating candidates is higher than ballots that contain only a subset of them, leading to \emph{Complete Ballots} being predicted much more often than sincere \emph{Take the X Best} behavior (See Table \ref{tab:approval-au}). 

Finally, we addressed this with a novel model of the attainability-utility heuristic with a threshold for approval voting. This model takes into account human cognitive constraints by generating an attainability-utility score for each candidate, rather than for every possible subset, and choosing to vote for only the candidates above a certain threshold. This model performed the best of all the models evaluated.  On average, our AUT model performed 80.7\% better than the AU model. For single-winner voting, it performed 57.5\% better than the second best model (\emph{Optimal}). In 2-winner elections, it performed 55.0\% better than the second best model (\emph{Take $k$ Best}). Finally, in the 3-winner elections, it performed 14.3\% better than the second best model (\emph{Take $k$ Best}).  Hence, our cognitively inspired heuristic model accurately predicts the voting behavior of agents in multi-winner approval voting environments more often than any existing models in the literature.

\section{Conclusion and Future Directions}

We have evaluated several heuristics as models of voter behavior in multi-winner approval voting. We found that our novel model of the attainability-utility heuristic with threshold provided the best predictive power of the models tested. This model simulates voters ranking each candidate based on a trade-off between attainability and utility, and then approving candidates ranked above a threshold. 

%Of special note, AU is a special case of AUT. In single-winner plurality elections where voters can only vote for one candidate, the AUT model simplifies to the AU model, accounting for human voting behavior across both plurality and approval voting scenarios *(i think this is true..jason)**. %. NOTE: Jason also says that our AUT model is better at prediction and even  -- Jason says we throw in a sentence in the discussion that points out that there may be other ways to do this which underscores the issues with generalizing the model.

To enhance prediction and understanding, a full taxonomy of (internal) cognitive strategies and capabilities along with (external) voting contexts and elements of uncertainty is required to fully explore the interaction between the two. Our AUT model, which describes the trade-off between attainability and utility at the candidate level while generalizing across different conditions of uncertainty, number of winners, and voting rules (i.e. approval and plurality), is an initial step in this direction. Going forward, we will explore how cognitively plausible models like AUT can be used to develop hybrid machine learning models that leverage models of cognitively plausible heuristics to predict voter behavior with even greater accuracy.

% Bibliography
\subsection*{Ethics Statement}
Our experiments on Mechanical Turk were conducted in an ethical manner and were overseen by the IRB Process at our University, IRB number will be added to the final version (to not break anonymity).  All our respondents were paid at least \$8.00 per hour of their time.

Our work is meant to advance the understanding of how people vote in realistic settings.  Given this, there is always a concern that detailed understanding of how people behave in high stakes decision making environments could be used to adverse ends.  However, a better understanding of the heuristics people naturally use may lead us to be able to detect deviations in the future and protect the integrity of election systems.

\subsection*{Acknowledgements}
Nicholas Mattei and K. Brent Venable are supported by NSF Award IIS-2007955.

{\small
\bibliography{abb,voting,heuristics}
}
\end{document}